\documentclass[11pt,a4paper]{article}

\usepackage{amsmath,amssymb,graphicx,siunitx,multirow,float,calc,array,tabularx,caption,float,subfig}
\usepackage[table]{xcolor}
\usepackage[colorlinks=true,breaklinks=true,allcolors=blue]{hyperref}
\usepackage[numbers,sort&compress]{natbib}
\usepackage{authblk}
\usepackage{mdframed}
\usepackage[left=1.5cm,right=1.5cm,bottom=1.5cm,top=2cm]{geometry}

\allowdisplaybreaks[4]

\begin{document}
\title{Total fraction of drug released from diffusion-controlled\\ delivery systems with binding reactions}
\author{Elliot J. Carr\thanks{\href{mailto:elliot.carr@qut.edu.au}{elliot.carr@qut.edu.au}}}
\affil{School of Mathematical Sciences, Queensland University of Technology, Brisbane, Australia.}

\date{}

\maketitle

\section*{Abstract} 
\noindent 
In diffusion-controlled drug delivery, it is possible for drug molecules to bind to the carrier material and never be released. A common way to incorporate this phenomenon into the governing mechanistic model is to include an irreversible first-order reaction term, where drug molecules become permanently immobilised once bound. For diffusion-only models, all the drug initially loaded into the device is released, while for reaction-diffusion models only a fraction of the drug is ultimately released. In this short paper, we show how to calculate this fraction for several common diffusion-controlled delivery systems. Easy-to-evaluate analytical expressions for the fraction of drug released are developed for monolithic and core-shell systems of slab, cylinder or sphere geometry. The developed formulas provide analytical insight into the effect that system parameters (e.g. diffusivity, binding rate, core radius) have on the total fraction of drug released, which may be helpful for practitioners designing drug delivery systems.

\medskip
\noindent\textit{Keywords}: drug delivery, binding, reaction-diffusion, core-shell, release profile.

\section{Introduction}
\begin{figure}[p]
\centering
\def\figh{0.35\textwidth}
\includegraphics[height=\figh]{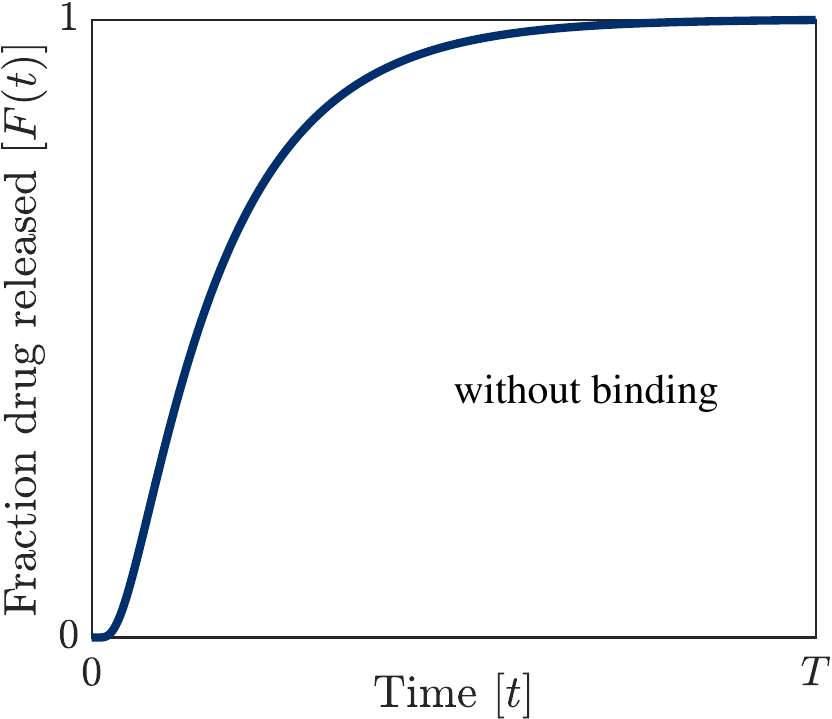}\hspace{1cm}\includegraphics[height=\figh]{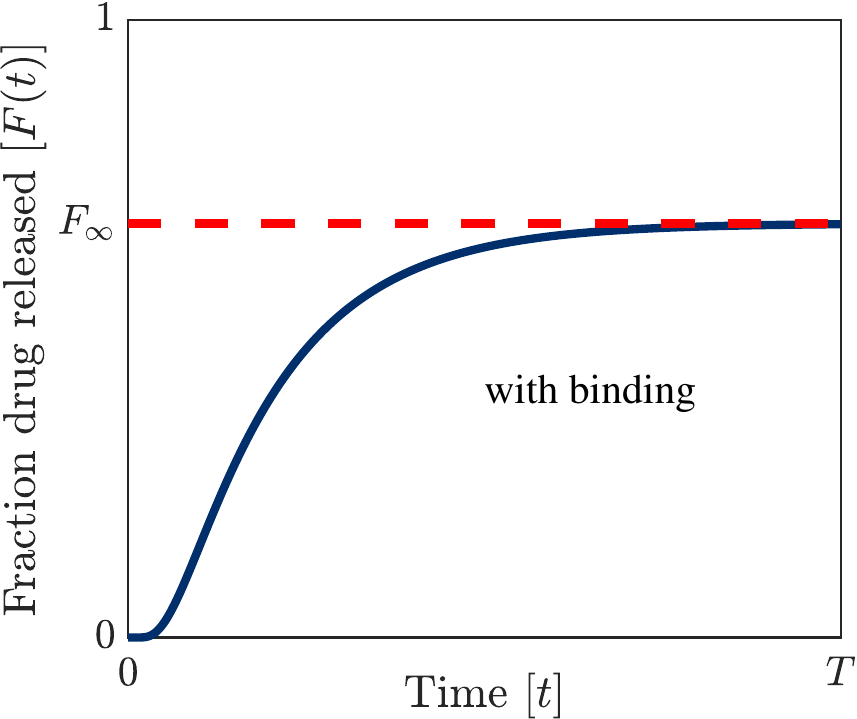}
\caption{Typical drug release profile, $F(t)$ (cumulative amount of drug released over the time interval $[0,t]$ divided by the initial amount of drug loaded into the device), without binding reactions and with binding reactions. In the long time limit, the fraction of drug released $F_{\infty} := \lim\limits_{t\rightarrow\infty}F(t)<1$ with binding.}
\label{fig:release_profiles}
\end{figure}

Increasingly, mechanistic mathematical models of drug delivery (based on physical conservation laws) are being developed to improve understanding of the transport mechanisms that control the release rate, explore the effect of varying design parameters on the release profile and avoid costly and time consuming experiments \cite{siepmann_2012}. In this field of research, mechanistic mathematical models of \emph{diffusion-controlled} drug delivery are typically based on Fick's second law, where the drug concentration within the system evolves in space and time according to the diffusion equation and specified initial and boundary conditions \cite{kaoui_2018,siepmann_2008,hadjitheodorou_2014,arifin_2006,carr_2018,simon_2019,ignacio_2021,gomes-filho_2020,carr_2022,peppas_2014,ritger_1987}. Such purely-diffusive models yield a release profile $F(t)$ (cumulative amount of drug released over the time interval $[0,t]$ divided by the initial amount of drug loaded into the device) that increases from zero initially to one in the long time limit (Figure \ref{fig:release_profiles}). Recent experimental work \cite{toniolo_2018}, however, has revealed that it is possible for drug to be trapped within the device and never be released due to drug molecules binding to the carrier material \cite{jain_2022,pontrelli_2021}. This reduces the amount of drug delivered, resulting in a release profile that approaches a value less than one, denoted in this paper by $F_{\infty}$, in the long time limit~(Figure \ref{fig:release_profiles}). Several recent mechanistic models have incorporated binding into the governing diffusion model using an irreversible first-order reaction term, where drug molecules become permanently immobilised once bound \cite{pontrelli_2021,jain_2022,carr_2024,mcginty_2015}.

\begin{figure}[p]
\centering
\def\figw{0.24\textwidth}
\includegraphics[width=\textwidth]{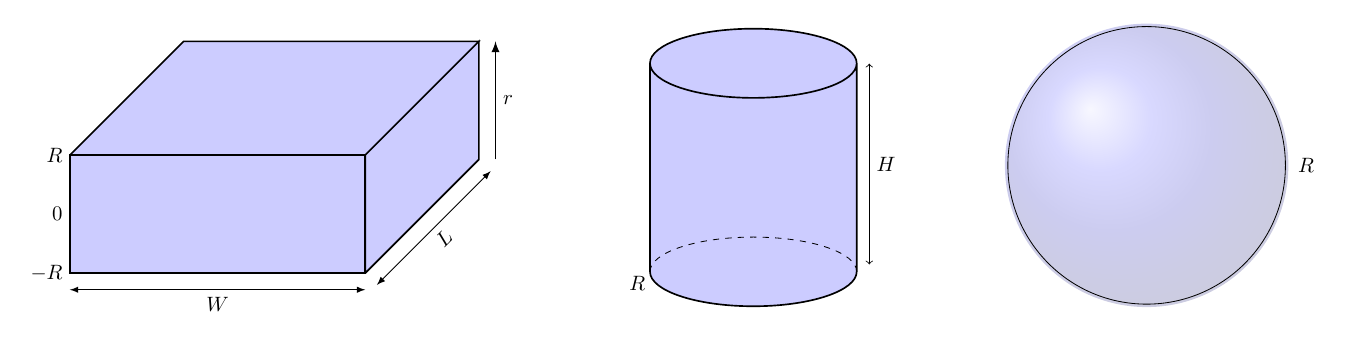}\\[-0.5cm]
\caption{\textit{Monolithic System.}~Slab, cylinder and sphere devices of radius $R$ consisting of a single homogeneous carrier material (with constant diffusivity). For the slab, $R$ is assumed to be significantly smaller than both $L$ and $W$ so that drug release is negligible through the minor surfaces (of area $2RL$ or $2RW$) compared to the two major surfaces of the slab (of area $LW$). For the cylinder, $R$ is significantly smaller than $H$ so that the drug release is negligible through the ends (of area $\pi R^{2}$) compared to the major surface of the cylinder (of area $2\pi RH$). The initial amount of drug is assumed to be homogeneously distributed throughout the device and drug molecules may bind to the carrier material anywhere within the device.}
\label{fig:monolithic_system}
\end{figure}

\begin{figure}[p]
\includegraphics[width=\textwidth]{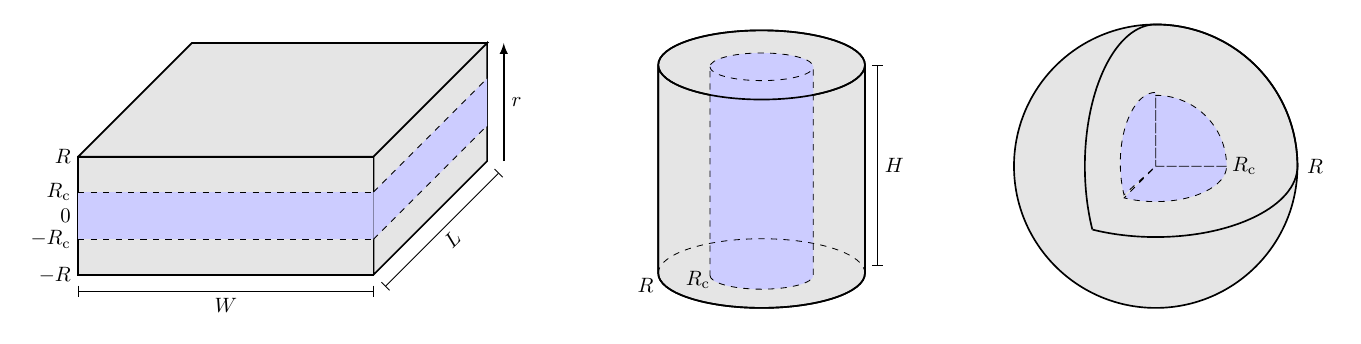}\\[-0.5cm]
\caption{\textit{Core-Shell System.}~Slab, cylinder and sphere devices of radius $R$ consisting of two homogeneous ``layers'' of carrier material (core of radius $R_{\mathrm{c}}$ encapsulated by a shell) with distinct diffusivities. For the slab, $R$ is assumed to be significantly smaller than both $L$ and $W$ so that drug release is negligible through the minor surfaces (of area $2RL$ or $2RW$) compared to the two major surfaces of the slab (of area $LW$). For the cylinder, $R$ is significantly smaller than $H$ so that the drug release is negligible through the ends (of area $\pi R^{2}$) compared to the major surface of the cylinder (of area $2\pi RH$). The initial amount of drug is assumed to be homogeneously distributed throughout the core only and drug molecules may bind to the carrier material within the shell only.}
\label{fig:core-shell_system}
\end{figure}

In this short paper, we develop a set of analytical expressions for $F_{\infty}$ for several  diffusion-controlled delivery systems with first-order binding reactions. Our study is valid under standard assumptions: the system is radially symmetric (drug concentration is a function of radius and time only), the boundary of the system is stationary (swelling or erosion of the device is negligible) and the drug is initially completely dissolved (drug dissolution is instantaneous). Compact and elegant expressions for $F_{\infty}$ are provided for both \textit{monolithic} and \textit{core-shell} systems of slab, cylinder or sphere geometry (Figures~\ref{fig:monolithic_system}--\ref{fig:core-shell_system}). The monolithic system consists of a single homogeneous carrier material (with constant diffusivity) while the core-shell system, consists of two homogeneous ``layers'' of carrier material (core and shell) with distinct diffusivities. For the monolithic system, the initial amount of drug is assumed to be homogeneously distributed throughout the system while for the core-shell system, the initial amount of drug is assumed to be homogeneously distributed throughout the core only. For the monolithic system, drug molecules may bind to the carrier material anywhere within the device while for the core-shell system, drug molecules may bind to the carrier material within the shell only \cite{pontrelli_2021}. Both the monolithic and core-shell systems are encapsulated in a thin coating (shell) that is either fully-permeable (no resistance to drug release) or semi-permeable (finite resistance to drug released) \cite{kaoui_2018}. In total, 12 distinct expressions for $F_{\infty}$ are derived, one for each combination of system type (monolithic, core-shell), device geometry (slab, cylinder, sphere) and coating permeability (fully-permeable, semi-permeable). Each expression explains how the value of $F_{\infty}$ changes for general values of the model parameters (e.g. diffusivity, reaction rate, surface transfer coefficient). 

The remaining sections of this paper discuss how the expressions for $F_{\infty}$ are developed for the monolithic and core-shell systems, respectively. Analytical and numerical evidence is then presented supporting the derived results.

\section{Monolithic System}
For the monolithic system, the drug dynamics are assumed to be governed by the reaction-diffusion model:
\begin{gather}
\label{eq:model_m_pde}
\frac{\partial c}{\partial t} = \frac{D}{r^{d-1}}\frac{\partial}{\partial r}\left(r^{d-1}\frac{\partial c}{\partial r}\right) - kc,\qquad 0 < r < R,\\
\label{eq:model_m_ic}
c(r,0) = c_{0},\\
\begin{cases} 
\text{$\dfrac{\partial c}{\partial r}(0,t) = 0,$} & \text{if $d=1,$}\\
\text{$c(0,t)$ finite}, & \text{if $d=2,3,$}\\
\end{cases}
\qquad
\label{eq:model_m_bc} 
\begin{cases} 
\text{$c(R,t) = 0,$} & \text{if fully-permeable},\\
\text{$-D\dfrac{\partial c}{\partial r}(R,t) = Pc(R,t)$}, & \text{if semi-permeable},
\end{cases}
\end{gather}
where $d$ specifies the geometry ($d=1,2,3$ for the slab, cylinder and sphere, respectively), $c(r,t)$ is the drug concentration, $R$ is the device radius (Figure \ref{fig:monolithic_system}), $D$ is the diffusivity, $k$ is the binding rate, $c_{0}$ is the initial uniform drug concentration and $P$ is the surface transfer coefficient. Note that the surface boundary condition (at $r = R$) depends on whether the coating is fully-permeable or semi-permeable with equivalence obtained in the limit as the surface transfer coefficient $P$ tends to infinity. Mathematically, the total fraction of drug released $F_{\infty}$ is given by the integral of the concentration flux over the release surface(s) divided by the initial amount of drug loaded into the device, which as shown in Appendix \ref{app:monolithic_initial} simplifies to 
\begin{gather}
\label{eq:Finf_m_og}
F_{\infty} = \frac{d}{Rc_{0}}\int_{0}^{\infty}-D\frac{\partial c}{\partial r}(R,t)\,\text{d}t,
\end{gather}
under radial symmetry. In Appendix \ref{app:monolithic}, we show further that $F_{\infty}$ can be expressed as
\begin{gather}
\label{eq:Finf_m}
F_{\infty} = 1-\frac{kd}{R^{d}c_{0}}\int_{0}^{R}r^{d-1}\mathcal{C}(r)\,\text{d}r,
\end{gather}
where $\mathcal{C}(r)$ satisfies the boundary value problem:
\begin{gather}
\label{eq:bvp_m_ode}
\frac{D}{r^{d-1}}\frac{\text{d}}{\text{d}r}\left(r^{d-1}\frac{\text{d}\mathcal{C}}{\text{d}r}\right) - k\mathcal{C} = -c_{0},\\
\label{eq:bvp_m_bc}
\begin{cases} 
\text{$\dfrac{\partial c}{\partial r}(0,t) = 0,$} & \text{if $d=1,$}\\
\text{$c(0,t)$ finite}, & \text{if $d=2,3,$}\\
\end{cases}
\qquad 
\begin{cases} 
\text{$c(R,t) = 0,$} & \text{if fully-permeable},\\
\text{$-D\dfrac{\partial c}{\partial r}(R,t) = Pc(R,t)$}, & \text{if semi-permeable}.\\
\end{cases}
\end{gather}
The attraction here is that the above boundary value problem admits exact analytical solutions involving hyperbolic ($d=1,3$) and Bessel functions ($d=2$) that can be found by hand or using a computer algebra system. Solving the boundary value problem (\ref{eq:bvp_m_ode})--(\ref{eq:bvp_m_bc}) separately for $d=1,2,3$, and evaluating the integral in (\ref{eq:Finf_m}) yields the expressions for $F_{\infty}$ given below in equations (\ref{eq:Finf_mslab1})--(\ref{eq:Finf_msphere2}). For $d = 2$, the expressions for $F_{\infty}$ involve $I_{0}$ and $I_{1}$, the modified Bessel functions of the first kind of zero and first order respectively.

\bigskip\medskip
\noindent\fbox{\begin{minipage}{0.98\textwidth}
\noindent\textbf{Fully-permeable coating}
\begin{alignat}{2}
\label{eq:Finf_mslab1}
&\textit{Slab} &\qquad F_{\infty} &= \frac{\sinh(\lambda)}{\lambda\cosh(\lambda )}\\[0.2cm]
&\textit{Cylinder} & F_{\infty} &= \frac{2I_{1}(\lambda)}{\lambda I_{0}(\lambda )}\\[0.2cm]
&\textit{Sphere} & F_{\infty} &= \frac{3[\lambda\cosh(\lambda) - \sinh(\lambda )]}{\lambda^{2}\sinh(\lambda)}
\end{alignat}
where $\lambda = \sqrt{\frac{kR^{2}}{D}}$.
\end{minipage}}

\bigskip\medskip
\noindent\fbox{\begin{minipage}{0.98\textwidth}
\noindent\textbf{Semi-permeable coating}
\begin{alignat}{2}
\label{eq:Finf_mslab2}
&\textit{Slab} &\qquad F_{\infty} &= \frac{\mathcal{P}\sinh(\lambda)}{\lambda[\lambda\sinh(\lambda) + \mathcal{P}\cosh(\lambda)]}\\[0.2cm]
&\textit{Cylinder} & F_{\infty} &= \frac{2\mathcal{P} I_{1}(\lambda)}{\lambda[\lambda I_{1}(\lambda) + \mathcal{P} I_{0}(\lambda)]}\\[0.2cm]
\label{eq:Finf_msphere2}
&\textit{Sphere} & F_{\infty} &= \frac{3\mathcal{P}[\lambda\cosh(\lambda)- \sinh(\lambda)]}{\lambda^{2}[\lambda\cosh(\lambda) + (\mathcal{P} - 1)\sinh(\lambda)]}
\end{alignat}
where $\lambda = \sqrt{\frac{kR^{2}}{D}}$ and $\mathcal{P} = \frac{PR}{D}$.
\end{minipage}}

\bigskip\medskip
\noindent The above results reveal how $F_{\infty}$ depends on a single dimensionless variable ($\lambda$) for the fully-permeable coating and two dimensionless variables ($\lambda$, $\mathcal{P}$) for the semi-permeable coating. Studying these expressions, we see that equivalence between the semi-permeable and permeable cases is obtained when $P\rightarrow\infty$. This fact together with the observation that the terms independent of $P$ in the denominators of equations (\ref{eq:Finf_mslab2})--(\ref{eq:Finf_msphere2}) are positive, means that $F_{\infty}$ is always smaller for the semi-permeable case than the corresponding fully-permeable case. This observation can be explained by the surface resistance resulting in drug molecules remaining in the device longer, increasing the likelihood that they bind to the carrier material and never be released. In summary, the above expressions provide the total fraction of drug released from the monolithic device (total amount of drug released divided by the initial amount of drug loaded into the device) for general values of the diffusivity ($D$), binding rate ($k$), device radius ($R$) and surface transfer coefficient ($P$). Similar expressions to those above appear in the classical text \cite{crank_1975} for the dual process of absorption into a cylinder/sphere from the surrounding medium.

Finally, we note that the actual amount of drug released is obtained by multiplying $F_{\infty}$ by the initial amount of drug loaded into the device:
\begin{gather*}
\begin{cases} 2Rc_{0}LW, & \text{if $d=1$},\\ 
\pi R^{2}c_{0}H, & \text{if $d=2$},
\\ 4\pi R^{3}c_{0}/3, & \text{if $d=3$},\end{cases}
\end{gather*}
where $L$, $W$ and $H$ are defined in Figure \ref{fig:monolithic_system}.

\section{Core-Shell System}
For the core-shell system, the drug dynamics are assumed to be governed by the reaction-diffusion model:
\begin{gather}
\label{eq:model_cs_pde1}
\frac{\partial c_{\mathrm{c}}}{\partial t} = \frac{D_{\mathrm{c}}}{r^{d-1}}\frac{\partial}{\partial r}\left(r^{d-1}\frac{\partial c_{\mathrm{c}}}{\partial r}\right),\qquad 0 < r < R_{\mathrm{c}},\\
\label{eq:model_cs_pde2}
\frac{\partial c_{\mathrm{s}}}{\partial t} = \frac{D_{\mathrm{s}}}{r^{d-1}}\frac{\partial}{\partial r}\left(r^{d-1}\frac{\partial c_{\mathrm{s}}}{\partial r}\right) - k_{\mathrm{s}}c_{\mathrm{s}},\qquad R_{\mathrm{c}} < r < R,\\
\label{eq:model_cs_ic}
c_{\mathrm{c}}(r,0) = c_{0},\qquad c_{\mathrm{s}}(r,0) = 0,\\
\label{eq:model_cs_int}
c_{\mathrm{c}}(R_{\mathrm{c}},t) = c_{\mathrm{s}}(R_{\mathrm{c}},t),\qquad D_{\mathrm{c}}\frac{\partial c_{\mathrm{c}}}{\partial r}(R_{\mathrm{c}},t) = D_{\mathrm{s}}\frac{\partial c_{\mathrm{s}}}{\partial r}(R_{\mathrm{c}},t),\\
\begin{cases} 
\text{$\dfrac{\partial c_{\mathrm{c}}}{\partial r}(0,t) = 0,$} & \text{if $d=1,$}\\
\text{$c_{\mathrm{c}}(0,t)$ finite}, & \text{if $d=2,3,$}\\
\end{cases}
\qquad 
\label{eq:model_cs_bc}
\begin{cases} 
\text{$c_{\mathrm{s}}(R,t) = 0,$} & \text{if fully-permeable},\\
\text{$-D_{\mathrm{s}}\dfrac{\partial c_{\mathrm{s}}}{\partial r}(R,t) = Pc_{\mathrm{s}}(R,t)$}, & \text{if semi-permeable},
\end{cases}
\end{gather}
where $d$ specifies the geometry ($d=1,2,3$ for the slab, cylinder and sphere, respectively), $c_{\mathrm{c}}(r,t)$ is the drug concentration in the core ($0 < r < R_{\mathrm{c}}$), $c_{\mathrm{s}}(r,t)$ is the drug concentration in the shell ($R_{\mathrm{c}}<r<R$), $D_{\mathrm{c}}$ is the diffusivity in the core, $D_{\mathrm{s}}$ is the diffusivity in the shell, $k_{\mathrm{s}}$ is the binding rate (shell only), $c_{0}$ is the initial uniform drug concentration (core only) and $P$ is the surface transfer coefficient. For the core-shell system, the total fraction of drug released is given by (see Appendix \ref{app:core-shell_initial}):
\begin{gather}
\label{eq:Finf_cs_og}
F_{\infty} = \frac{dR^{d-1}}{R_{\mathrm{c}}^{d}c_{0}}\int_{0}^{\infty}-D_{\mathrm{s}}\frac{\partial c_{\mathrm{s}}}{\partial r}(R,t)\,\text{d}t.
\end{gather}
In Appendix \ref{app:core-shell}, we show that $F_{\infty}$ can be expressed as:
\begin{gather}
\label{eq:Finf_cs}
F_{\infty} = 1-\frac{k_{\mathrm{s}}d}{R_{\mathrm{c}}^{d}c_{0}}\int_{R_{\mathrm{c}}}^{R}r^{d-1}\mathcal{C}_{\mathrm{s}}(r)\,\text{d}r,
\end{gather}
where $\mathcal{C}_{\mathrm{s}}(r)$ satisfies the following boundary value problem:
\begin{gather}
\label{eq:bvp_cs_ode1}
\frac{D_{\mathrm{c}}}{r^{d-1}}\frac{\text{d}}{\text{d}r}\left(r^{d-1}\frac{\text{d}\mathcal{C}_{\mathrm{c}}}{\text{d}r}\right) = -c_{0},\qquad 0 < r < R_{\mathrm{c}},\\
\label{eq:bvp_cs_ode2}
\frac{D_{\mathrm{s}}}{r^{d-1}}\frac{\text{d}}{\text{d}r}\left(r^{d-1}\frac{\text{d}\mathcal{C}_{\mathrm{s}}}{\text{d}r}\right) - k_{\mathrm{s}}\mathcal{C}_{\mathrm{s}} = 0,\qquad R_{\mathrm{c}} < r < R,\\
\label{eq:bvp_cs_int}
\mathcal{C}_{\mathrm{c}}(R_{\mathrm{c}}) = \mathcal{C}_{\mathrm{s}}(R_{\mathrm{c}}),\qquad D_{\mathrm{c}}\frac{\text{d}\mathcal{C}_{\mathrm{c}}}{\text{d}r}(R_{\mathrm{c}}) = D_{\mathrm{s}}\frac{\text{d}\mathcal{C}_{\mathrm{s}}}{\text{d}r}(R_{\mathrm{c}}),\\
\label{eq:bvp_cs_bc}
\begin{cases} 
\text{$\dfrac{\text{d}\mathcal{C}_{\mathrm{c}}}{\text{d}r}(0) = 0,$} & \text{if $d=1,$}\\
\text{$\mathcal{C}_{\mathrm{c}}(0)$ finite}, & \text{if $d=2,3,$}\\
\end{cases}
\qquad 
\begin{cases} 
\text{$\mathcal{C}_{\mathrm{s}}(R) = 0,$} & \text{if fully-permeable},\\
\text{$-D_{\mathrm{s}}\dfrac{\text{d}\mathcal{C}_{\mathrm{s}}}{\text{d} r}(R) = P\mathcal{C}_{\mathrm{s}}(R)$}, & \text{if semi-permeable}.
\end{cases}
\end{gather}
As for the monolithic system, the above boundary value problem admits exact analytical solutions involving hyperbolic ($d=1,3$) and Bessel functions ($d=2$) that can be found by hand (quite tediously) or using a computer algebra system. Solving the boundary value problem (\ref{eq:bvp_cs_ode1})--(\ref{eq:bvp_cs_bc}) separately for $d=1,2,3$ and evaluating the integral in (\ref{eq:Finf_cs}) yields the expressions for $F_{\infty}$ given below in equations (\ref{eq:Finf_cs1})--(\ref{eq:Finf_cs2}). For $d = 2$, the expressions for $F_{\infty}$ involve $I_{0}$ and $I_{1}$, the modified Bessel functions of the first kind of zero and first order respectively and $K_{0}$ and $K_{1}$, the modified Bessel functions of the second kind of zero and first order respectively.

\bigskip\medskip
\noindent\fbox{\begin{minipage}{0.98\textwidth}
\noindent\textbf{Fully-permeable coating}
\begin{alignat}{2}
\label{eq:Finf_cs1}
&\hspace*{-0.3cm}\textit{Slab} &\qquad F_{\infty} &= \frac{1}{\cosh(\lambda_{\mathrm{s}}\mathcal{R}_{\mathrm{s}})}\\[0.2cm]
&\hspace*{-0.3cm}\textit{Cylinder} & F_{\infty} &= \frac{1}{\mathcal{R}_{\mathrm{c}}\lambda_{\mathrm{s}}[K_{0}(\lambda_{\mathrm{s}})I_{1}(\mathcal{R}_{\mathrm{c}}\lambda_{\mathrm{s}}) + I_{0}(\lambda_{\mathrm{s}})K_{1}(\mathcal{R}_{\mathrm{c}}\lambda_{\mathrm{s}})]}\\[0.2cm]
&\hspace*{-0.3cm}\textit{Sphere} & F_{\infty} &= \frac{\lambda_{\mathrm{s}}}{\sinh(\lambda_{\mathrm{s}}\mathcal{R}_{\mathrm{s}})+\lambda_{\mathrm{s}}\mathcal{R}_{\mathrm{c}}\cosh(\lambda_{\mathrm{s}}\mathcal{R}_{\mathrm{s}})}
\end{alignat}
where $\lambda_{\mathrm{s}} = \sqrt{\frac{k_{\mathrm{s}}R^{2}}{D_{\mathrm{s}}}}$, and $\mathcal{R}_{\mathrm{c}} = \frac{R_{\mathrm{c}}}{R}$.
\end{minipage}}

\bigskip\medskip
\noindent\fbox{\begin{minipage}{0.98\textwidth}
\noindent\textbf{Semi-permeable coating}
\begin{alignat}{2}
&\hspace*{-0.3cm}\textit{Slab} &\qquad F_{\infty} &= \frac{\mathcal{P}_{\mathrm{s}}}{\mathcal{P}_{\mathrm{s}}\cosh(\lambda_{\mathrm{s}}\mathcal{R}_{\mathrm{s}})+\lambda_{\mathrm{s}}\sinh(\lambda_{\mathrm{s}}\mathcal{R}_{\mathrm{s}})}\\[0.2cm]
&\hspace*{-0.3cm}\textit{Cylinder} & F_{\infty} &= \frac{\mathcal{P}_{\mathrm{s}}}{\mathcal{R}_{\mathrm{c}}\lambda_{\mathrm{s}}([\mathcal{P}_{\mathrm{s}}K_{0}(\lambda_{\mathrm{s}}) -\lambda_{\mathrm{s}}K_{1}(\lambda_{\mathrm{s}})]I_{1}(\mathcal{R}_{\mathrm{c}}\lambda_{\mathrm{s}}) + [\mathcal{P}_{\mathrm{s}}I_{0}(\lambda_{\mathrm{s}})+\lambda_{\mathrm{s}}I_{1}(\lambda_{\mathrm{s}})]K_{1}(\mathcal{R}_{\mathrm{c}}\lambda_{\mathrm{s}}))}\\[0.2cm]
&\hspace*{-0.3cm}\textit{Sphere} & F_{\infty} &= \frac{\mathcal{P}_{\mathrm{s}}\lambda_{\mathrm{s}}}{(\mathcal{P}_{\mathrm{s}}-1+\mathcal{R}_{\mathrm{c}}\lambda_{\mathrm{s}}^{2})\sinh(\lambda_{\mathrm{s}}\mathcal{R}_{\mathrm{s}})+\lambda_{\mathrm{s}}(\mathcal{R}_{\mathrm{s}}+\mathcal{P}_{\mathrm{s}}\mathcal{R}_{\mathrm{c}})\cosh(\lambda_{\mathrm{s}}\mathcal{R}_{\mathrm{s}})}
\label{eq:Finf_cs2}
\end{alignat}
where $\lambda_{\mathrm{s}} = \sqrt{\frac{k_{\mathrm{s}}R^{2}}{D_{\mathrm{s}}}}$, $\mathcal{P}_{\mathrm{s}} = \frac{PR}{D_{\mathrm{s}}}$, $\mathcal{R}_{\mathrm{c}} = \frac{R_{\mathrm{c}}}{R}$ and $\mathcal{R}_{\mathrm{s}} = 1 - \mathcal{R}_{\mathrm{c}}$.
\end{minipage}}

\bigskip\medskip
\noindent The above results reveal how $F_{\infty}$ depends on two dimensionless variables ($\lambda_{\mathrm{s}}$, $\mathcal{R}_{\mathrm{c}}$) for the fully-permeable coating and three dimensionless variables ($\lambda_{\mathrm{s}}$, $\mathcal{R}_{\mathrm{c}}$ and $\mathcal{P}_{\mathrm{s}}$) for the semi-permeable coating. Interestingly (perhaps unexpectedly), $F_{\infty}$ is independent of the diffusivity in the core $D_{\mathrm{c}}$ in all cases. As for the monolithic system, $F_{\infty}$ is always smaller for the semi-permeable case with equivalence obtained when $P\rightarrow\infty$. In summary, the above expressions provide the total fraction of drug released from the core-shell device (total amount of drug released divided by the initial amount of drug loaded into the device) for general values of the core diffusivity ($D_{\mathrm{c}})$, shell diffusivity ($D_{\mathrm{s}}$), shell binding rate ($k_{\mathrm{s}}$), core radius ($R_{\mathrm{c}}$), device radius ($R$) and surface transfer coefficient ($P$). 

Finally, we note that the actual amount of drug released is obtained by multiplying $F_{\infty}$ by the initial amount of drug loaded into the device:
\begin{gather*}
\begin{cases} 2R_{\mathrm{c}}c_{0}LW, & \text{if $d=1$},\\ \pi R_{\mathrm{c}}^{2}c_{0}H, & \text{if $d=2$},
\\ 4\pi R_{\mathrm{c}}^{3}c_{0}/3, & \text{if $d=3$},\end{cases}
\end{gather*}
where $L$, $W$ and $H$ are defined in Figure \ref{fig:core-shell_system}.

\section{Supporting Numerical Experiments}
\begin{figure}[p]
\centering
\def\figw{0.3\textwidth}
(a) Fully-permeable coating\\[0.05cm]
\includegraphics[width=\figw]{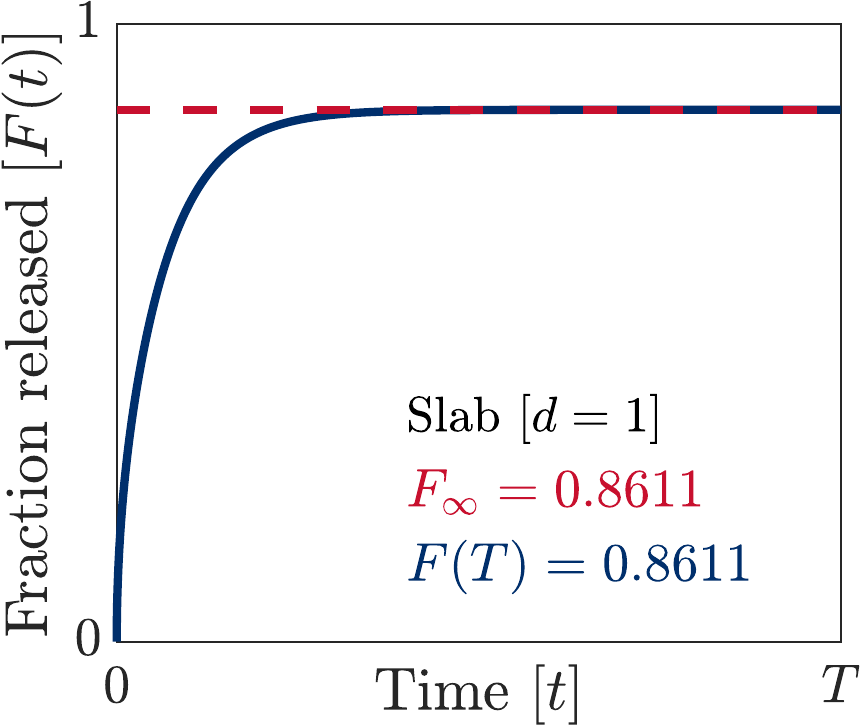}\hfill\includegraphics[width=\figw]{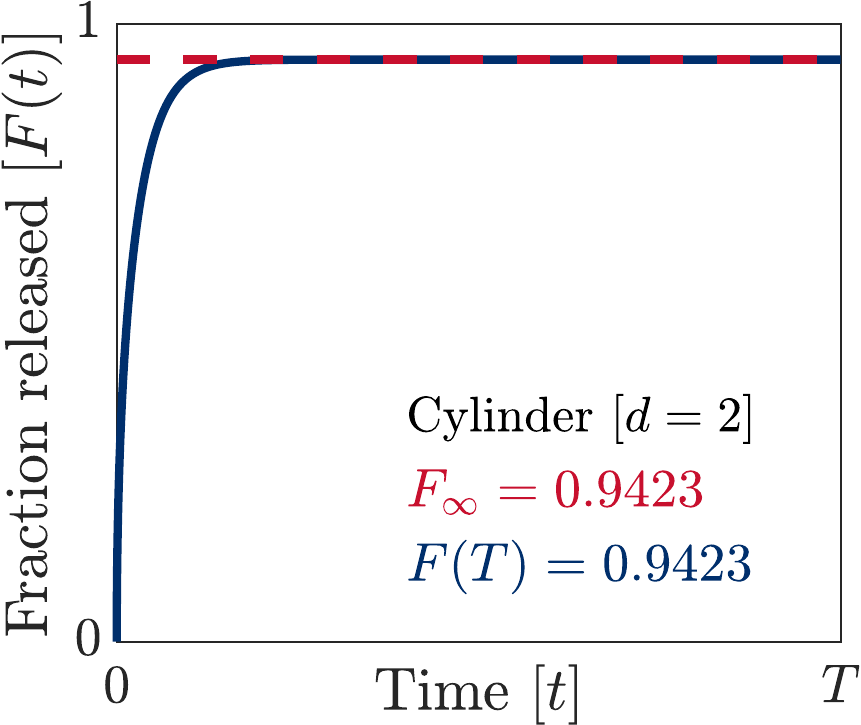}\hfill\includegraphics[width=\figw]{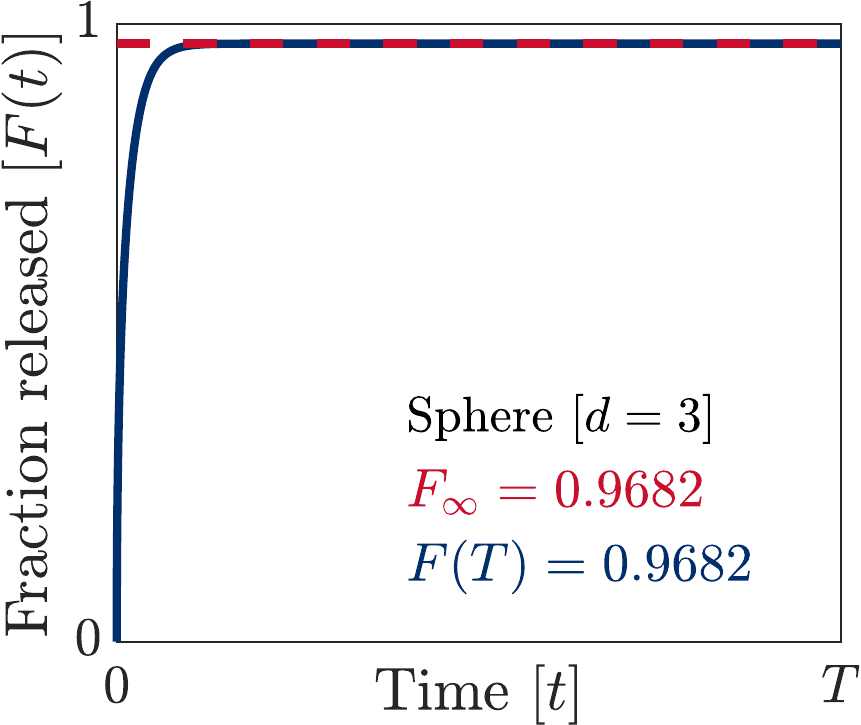}\\[0.2cm]
(b) Semi-permeable coating\\[0.05cm]
\includegraphics[width=\figw]{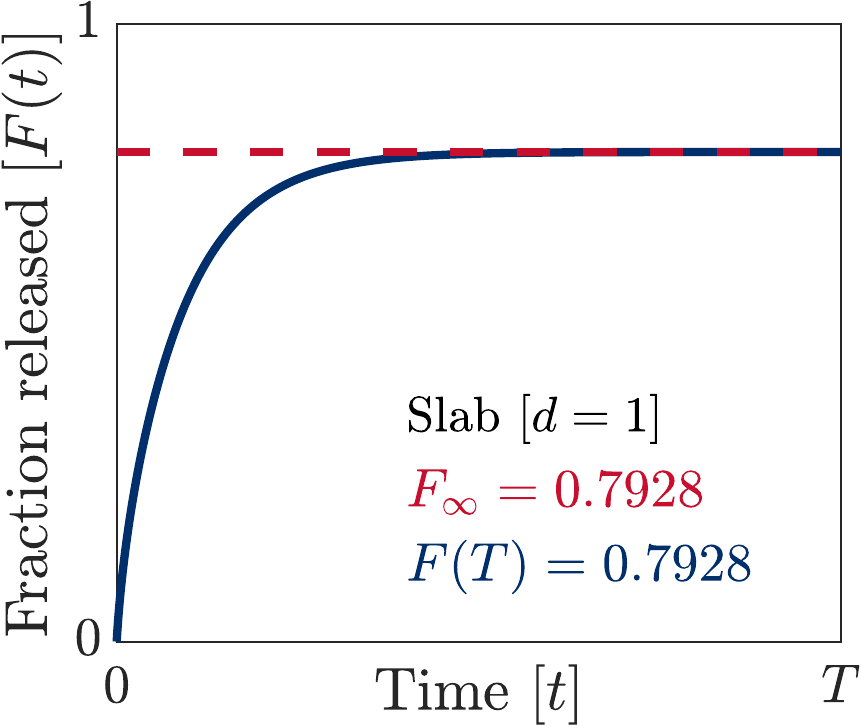}\hfill\includegraphics[width=\figw]{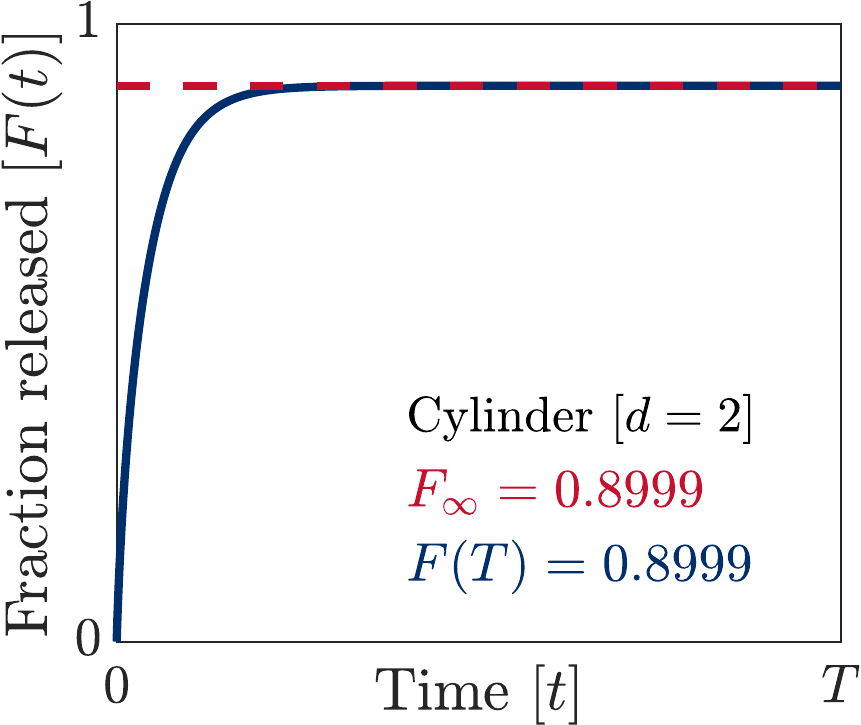}\hfill\includegraphics[width=\figw]{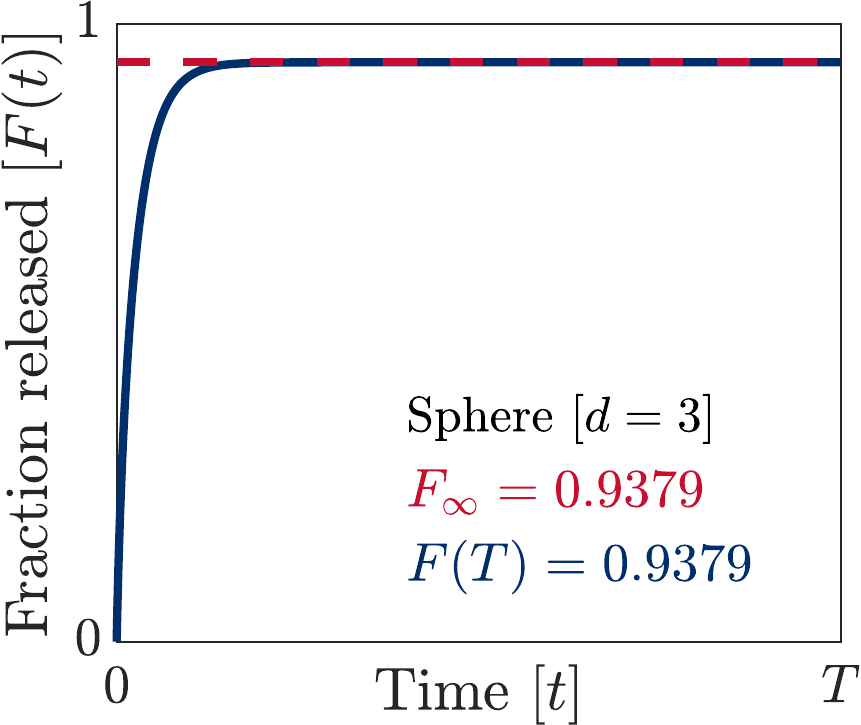}
\caption{\textit{Monolithic System.} Comparing the analytical expressions for $F_{\infty}$ (\ref{eq:Finf_mslab1})--(\ref{eq:Finf_msphere2}) to numerical approximations of $F(T)$, where $T$ is sufficiently large to capture the long time limiting behaviour of $F(t)$. Parameter values: $c_{0} = 0.4\,\text{mol}\,\text{cm}^{-3}$, $R = 10^{-4}\,\text{cm}$, $D = 10^{-12}\,\text{cm}^{2}\,\text{s}^{-1}$, $k = 5\times 10^{-5}\,\text{s}^{-1}$, $T = 5\times 10^{4}\,\text{s}$ (dimensional), $\lambda = \sqrt{0.5}$, $\mathcal{P} = 5$ (dimensionless).}
\label{fig:monolithic_results}

\centering
\def\figw{0.3\textwidth}
(a) Fully-permeable coating\\[0.05cm]
\includegraphics[width=\figw]{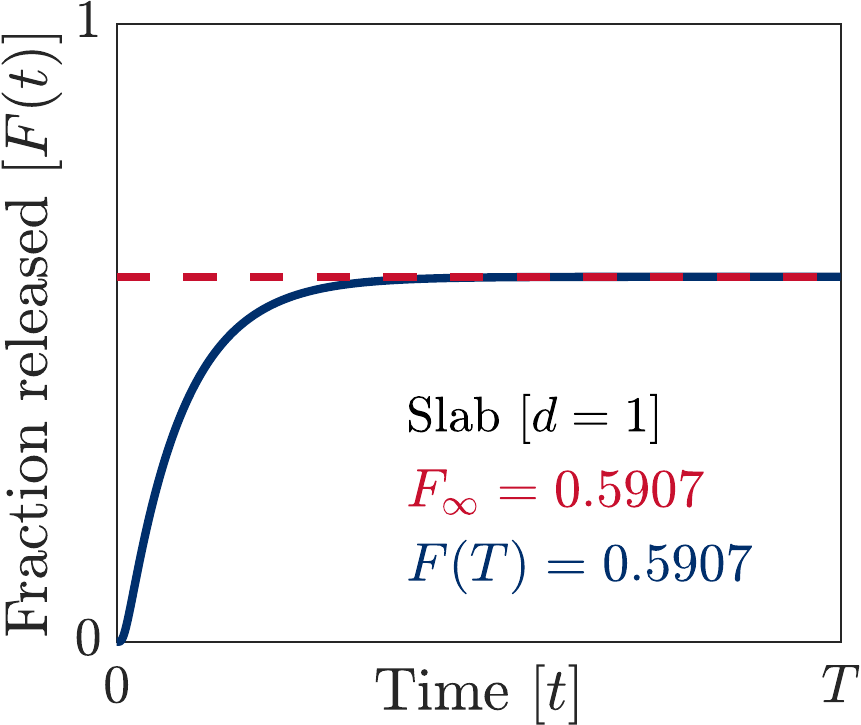}\hfill\includegraphics[width=\figw]{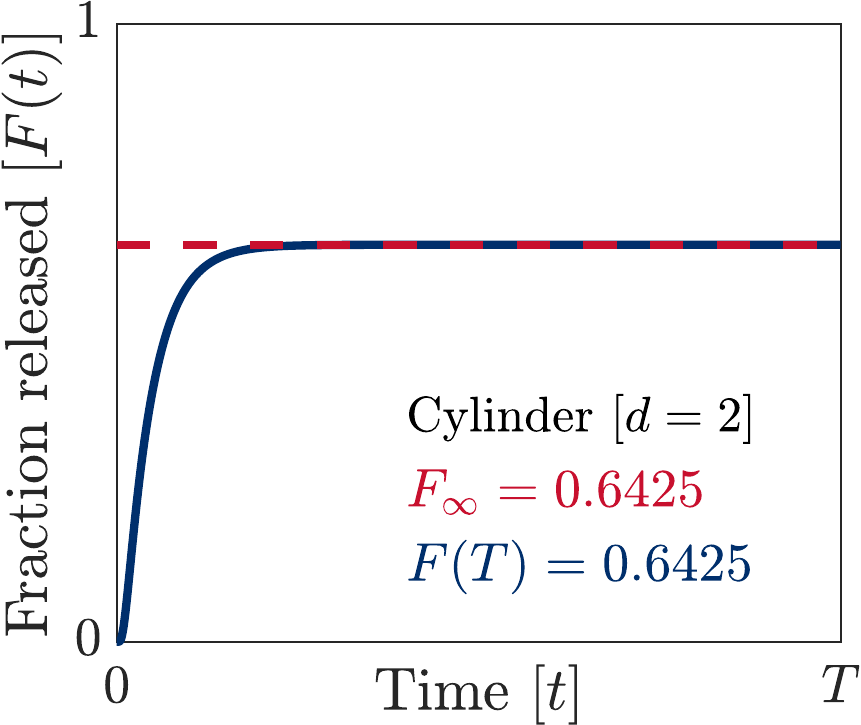}\hfill\includegraphics[width=\figw]{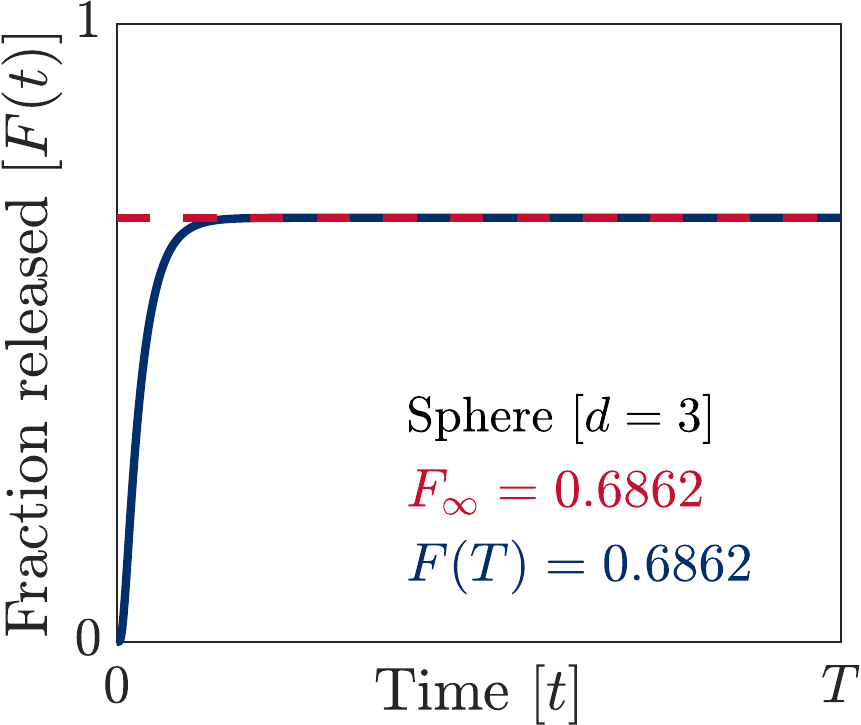}\\[0.2cm]
(b) Semi-permeable coating\\[0.05cm]
\includegraphics[width=\figw]{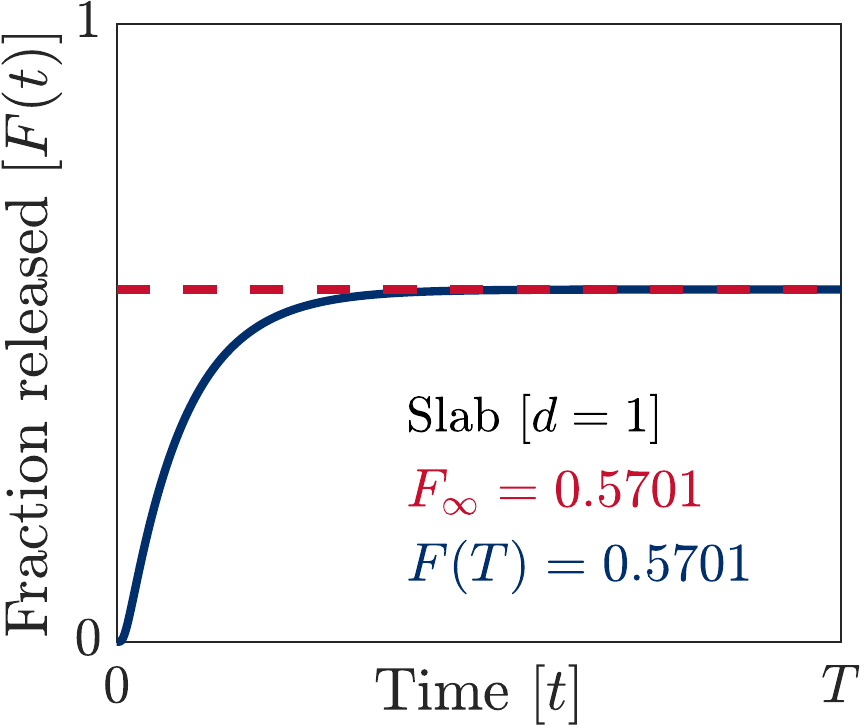}\hfill\includegraphics[width=\figw]{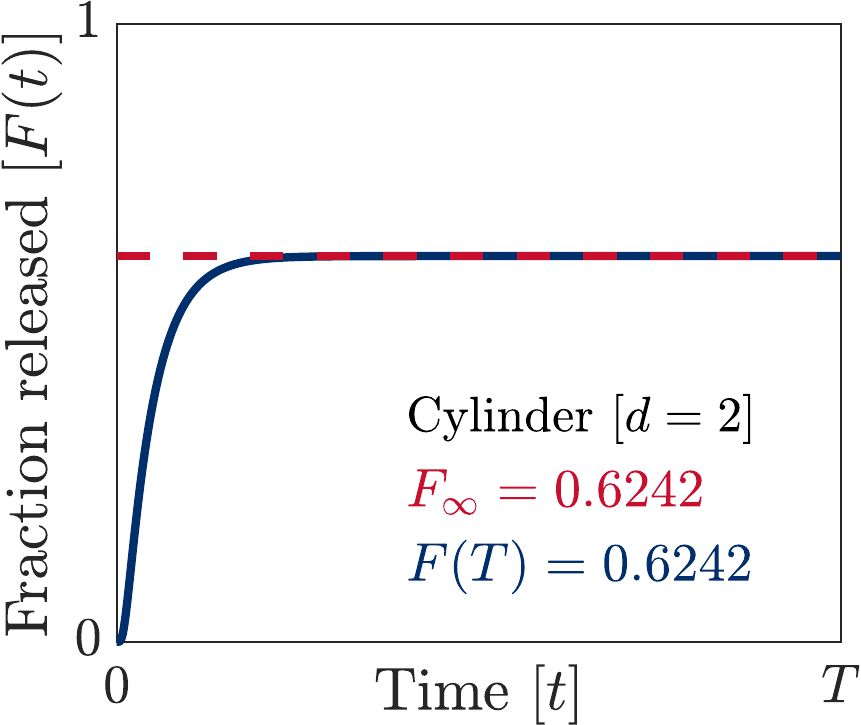}\hfill\includegraphics[width=\figw]{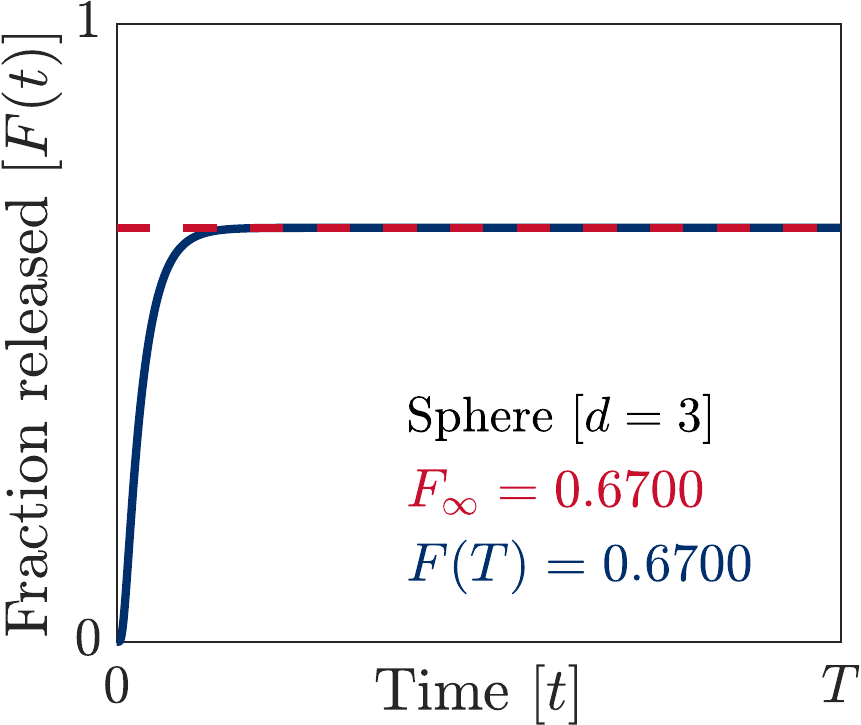}
\caption{\textit{Core-Shell System.} Comparing the analytical expressions for $F_{\infty}$ (\ref{eq:Finf_cs1})--(\ref{eq:Finf_cs2}) to numerical approximations of $F(T)$, where $T$ is sufficiently large to capture the long time limiting behaviour of $F(t)$. Parameter values: $c_{0} = 0.4\,\text{mol}\,\text{cm}^{-3}$, $R_{\mathrm{c}} = 0.5\times 10^{-4}\,\text{cm}$, $R = 10^{-4}\,\text{cm}$, $D_{\mathrm{c}} = 10^{-11}\,\text{cm}^{2}\,\text{s}^{-1}$, $D_{\mathrm{s}} = 10^{-13}\,\text{cm}^{2}\,\text{s}^{-1}$, $k_{\mathrm{s}} = 5\times 10^{-5}\,\text{s}^{-1}$, $T = 3\times 10^{5}\,\text{s}$ (dimensional), $\lambda_{\mathrm{s}} = \sqrt{5}$, $\mathcal{P}_{\mathrm{s}} = 50$, $\mathcal{R}_{\mathrm{c}} = 0.5$ (dimensionless).}
\label{fig:core-shell_results}
\end{figure}

The analytical expressions for $F_{\infty}$ derived in the previous sections have been verified analytically using MATLAB's symbolic toolbox. Results in Figures \ref{fig:monolithic_results} and \ref{fig:core-shell_results} also provide evidence to support the derived expressions for a set of parameter values \cite{carr_2024}. These figures compare computed values of $F_{\infty}$ to the long time limiting behaviour of the cumulative fraction of drug released: 
\begin{gather}
\label{eq:Ft}
F(t) = \begin{cases} \displaystyle\frac{d}{Rc_{0}}\int_{0}^{t}-D\frac{\partial c}{\partial r}(R,\tau)\,\text{d}\tau, & \text{if monolithic system},\\[0.4cm]
\displaystyle\frac{dR^{d-1}}{R_{\mathrm{c}}^{d}c_{0}}\int_{0}^{t}-D_{\mathrm{s}}\frac{\partial c_{\mathrm{s}}}{\partial r}(R,\tau)\,\text{d}\tau, & \text{if core-shell system}.
\end{cases}
\end{gather}
To compute $F(t)$ we solve the governing reaction-diffusion model, equations (\ref{eq:model_m_pde})--(\ref{eq:model_m_bc}) for the monolithic system and (\ref{eq:model_cs_pde1})--(\ref{eq:model_cs_bc}) for the core-shell system, numerically by (i) discretising over space $[0,R]$ using a finite volume method with $N_{r}$ uniformly spaced nodes and (ii) discretising over time $[0,T$] using the backward Euler method with $N_{t}$ time steps of fixed duration. Here $T$ is chosen to be sufficiently large to capture the long time limiting behaviour of $F(t)$. The numerical method computes discrete approximations to the concentration $c(r_{k}, t_{i})$ for $k = 1,\hdots,N_{r}$ and $i = 1,\hdots,N_{t}$, where $r_{k} = (k - 1)R/(N_{r} - 1)$ and $t_{i} = iT/N_{t}$. The numerical approximations to $c(r_{k}, t_{i})$ for $k = 1,\hdots,N_{r}$ are then used to compute $F(t_{i})$ for $i = 1,\hdots,N_{t}$ by applying a backward difference approximation to the radial concentration gradient in (\ref{eq:Ft}) followed by a trapezoidal rule approximation to the integral over $[0,t_{i}]$ in (\ref{eq:Ft}). Comparisons in Figures \ref{fig:monolithic_results} and \ref{fig:core-shell_results} are given for $N_{r} = 2001$ spatial nodes and $N_{t} = 10^{4}$ time steps with excellent agreement reported between the computed values of $F_{\infty}$ and $F(T)$ across all combination of system type (monolithic, core-shell), device geometry (slab, cylinder, sphere) and coating permeability (fully-permeable, semi-permeable). Full details on both the analytical and numerical verification are available in our MATLAB code, which is available on GitHub at \href{https://github.com/elliotcarr/Carr2024a}{https://github.com/elliotcarr/Carr2024a}. 

\section{Conclusions}
In summary, we have developed a set of analytical expressions to calculate $F_{\infty}$ (total amount of drug released divided by the initial amount of drug) for several diffusion-controlled delivery systems in the presence of binding reactions. In total, 12 distinct expressions for $F_{\infty}$ were derived, one for each combination of system type (monolithic, core-shell), device geometry (slab, cylinder, sphere) and coating permeability (fully-permeable, semi-permeable). Each expression provides analytical insight into the effect that particular physical and geometrical parameters (device geometry, diffusivity, binding rate, surface transfer coefficient) have on the value of $F_{\infty}$; insight which may be helpful for practitioners designing drug delivery systems.

While we have considered a wide-range of delivery systems in this work, it is important to note that the results are limited to the specific device configurations presented. Extending the analysis for the core-shell system to explore, for example, binding reactions in the core and/or non-zero initial drug concentration in the shell would yield different expressions for $F_{\infty}$.

\appendix
\section{Appendices}

\subsection{Monolithic System: Initial form for $F_{\infty}$}
\label{app:monolithic_initial}
The total fraction of drug released, defined as the integral of the concentration flux through the release surface(s) ($\omega$) scaled by the initial amount of drug loaded into the device region ($\Omega$), simplifies to (\ref{eq:Finf_m_og}) as follows
\begin{align*}
F_{\infty} &= \frac{\int_{0}^{\infty}\iint_{\omega}\left[-D\nabla c\cdot\mathbf{n}\,\text{d}A\right]\,\text{d}t}{\iiint_{\Omega} c_{0}\,\text{d}V} = \frac{|\omega|\int_{0}^{\infty}-D\frac{\partial c}{\partial r}(R,t)\,\text{d}t}{|\Omega| c_{0}} = \frac{d}{Rc_{0}}\int_{0}^{\infty}-D\frac{\partial c}{\partial r}(R,t)\,\text{d}t,
\end{align*}
due to radial symmetry and the fact that $|\omega|/|\Omega|$ equals the surface area of the ($d-1$)-dimensional sphere of radius $R$ divided by the volume of the $d$-dimensional ball of radius $R$ ($LW$ and $H$ cancel from both the numerator and denominator for the slab and cylinder, respectively).

\subsection{Monolithic System: Alternative form for $F_{\infty}$}
\label{app:monolithic}
In this appendix, we demonstrate equivalence of the two forms of $F_{\infty}$ given in equations (\ref{eq:Finf_m_og}) and (\ref{eq:Finf_m}). Integrating the reaction-diffusion equation (\ref{eq:model_m_pde}) over the slab, cylinder and sphere yields
\begin{gather*}
\int_{0}^{R}r^{d-1}\frac{\partial c}{\partial t}\,\text{d}r = D\int_{0}^{R}\frac{\partial}{\partial r}\left(r^{d-1}\frac{\partial c_{\mathrm{c}}}{\partial r}\right)\,\text{d}r - k\int_{0}^{R}r^{d-1}c\,\text{d}r,
\end{gather*}
due to radial symmetry. Reversing the order of integration and differentiation on the left and evaluating the first integral on the right produces:
\begin{gather*}
\frac{\text{d}}{\text{d}t}\left[\int_{0}^{R}r^{d-1} c\,\text{d}r\right] = R^{d-1}\left[D\frac{\partial c}{\partial r}(R,t)\right] - k\int_{0}^{R}r^{d-1}c\,\text{d}r,
\end{gather*}
and hence:
\begin{gather*}
R^{d-1}\left[-D\frac{\partial c}{\partial r}(R,t)\right] = -\frac{\text{d}}{\text{d}t}\left[\int_{0}^{R}r^{d-1} c\,\text{d}r\right] - k\int_{0}^{R}r^{d-1}c\,\text{d}r.
\end{gather*}
Integrating this last equation over the time interval $[0,\infty)$, using the initial condition (\ref{eq:model_m_ic}) and the fact that $c(r,t)$ tends to zero at infinite time yields:
\begin{gather*}
R^{d-1}\int_{0}^{\infty}\left[-D\frac{\partial c}{\partial r}(R,t)\right]\,\text{d}t = \int_{0}^{R}r^{d-1}c_{0}\,\text{d}r - k\int_{0}^{R}r^{d-1}\mathcal{C}(r)\,\text{d}r,
\end{gather*}
where $\mathcal{C}(r) := \int_{0}^{\infty} c(r,t)\,\text{d}t$. Dividing both sides of this last equation by $\int_{0}^{R}r^{d-1}c_{0}\,\text{d}r = R^{d}c_{0}/d$ then gives the desired result:
\begin{gather*}
\frac{d}{Rc_{0}}\int_{0}^{\infty}-D\frac{\partial c}{\partial r}(R,t)\,\text{d}t = 1-\frac{kd}{R^{d}c_{0}}\int_{0}^{R}r^{d-1}\mathcal{C}(r)\,\text{d}r.
\end{gather*}
To complete the demonstration, we need to show that $\mathcal{C}(r)$ satisfies the boundary value problem (\ref{eq:bvp_m_ode})--(\ref{eq:bvp_m_bc}). The differential equation (\ref{eq:bvp_m_ode}) is obtained by integrating the reaction-diffusion equation (\ref{eq:model_m_pde}) over the time interval $[0,\infty)$, using the initial condition (\ref{eq:model_m_ic}) and the fact that $c(r,t)$ tends to zero at infinite time:
\begin{gather*}
-c_{0} = \int_{0}^{\infty}\frac{D}{r^{d-1}}\frac{\partial}{\partial r}\left(r^{d-1}\frac{\partial c}{\partial r}\right)\,\text{d}t - k\int_{0}^{\infty}c(r,t)\,\text{d}t,
\end{gather*}
which simplifies to equation (\ref{eq:bvp_m_ode}) when reversing the order of differentiation and integration in the second term on the right and introducing $C(r) := \int_{0}^{\infty} c(r,t)\,\text{d}t$. Finally, the boundary conditions (\ref{eq:bvp_m_bc}) are obtained by integrating the boundary conditions (\ref{eq:model_m_bc}) over the time interval $[0,\infty)$:
\begin{gather*}
\begin{cases} \displaystyle\int_{0}^{\infty}\frac{\partial c}{\partial r}(0,t)\,\text{d}t = 0, & \text{if $d=1$},\\[0.4cm]
\text{$\displaystyle\int_{0}^{\infty}c(0,t)\,\text{d}t$ finite}, & \text{if $d=2,3$},
\end{cases}
\qquad
\begin{cases} \displaystyle\int_{0}^{\infty} c(R,t)\,\text{d}t = 0, & \text{if fully-permeable},\\[0.4cm]
\displaystyle-D\int_{0}^{\infty}\frac{\partial c}{\partial r}(R,t)\,\text{d}t = P\int_{0}^{\infty} c(R,t)\,\text{d}t, & \text{if semi-permeable},\\
\end{cases}
\end{gather*}
and reversing the order of differentiation and integration as appropriate.

\subsection{Core-Shell System: Initial form for $F_{\infty}$}
\label{app:core-shell_initial}
The total fraction of drug released, defined as the integral of the concentration flux through the release surface(s) ($\omega$) scaled by the initial amount of drug loaded into the core region ($\Omega_{\mathrm{c}}$), simplifies to (\ref{eq:Finf_cs_og}) as follows:
\begin{align*}
F_{\infty} &= \frac{\int_{0}^{\infty}\left[\int_{\omega}-D_{\mathrm{s}}\nabla c_{\mathrm{s}}\cdot\mathbf{n}\,\text{d}A\right]\,\text{d}t}{\int_{\Omega_{\mathrm{c}}} c_{0}\,\text{d}V} = \frac{|\omega|\int_{0}^{\infty}-D\frac{\partial c}{\partial r}(R,t)\,\text{d}t}{|\Omega_{\mathrm{c}}| c_{0}} = \frac{dR^{d-1}}{R_{\mathrm{c}}^{d}c_{0}}\int_{0}^{\infty}-D_{\mathrm{s}}\frac{\partial c_{\mathrm{s}}}{\partial r}(R,t)\,\text{d}t,
\end{align*}
due to radial symmetry and the fact that $|\omega|/|\Omega_{\mathrm{c}}|$ equals the surface area of the ($d-1$)-dimensional sphere of radius $R$ divided by the volume of the $d$-dimensional ball of radius $R_{\mathrm{c}}$ ($LW$ and $H$ cancel from both the numerator and denominator for the slab and cylinder, respectively).

\subsection{Core-Shell System: Alternative form for $F_{\infty}$}
\label{app:core-shell}
In this appendix, we demonstrate equivalence of the two forms of $F_{\infty}$ given in equations (\ref{eq:Finf_cs_og}) and (\ref{eq:Finf_cs}). Integrating the reaction-diffusion equation (\ref{eq:model_cs_pde1})--(\ref{eq:model_cs_pde2}) over the slab, cylinder and sphere yields
\begin{gather*}
\int_{0}^{R_{\mathrm{c}}}r^{d-1}\frac{\partial c_{\mathrm{c}}}{\partial t}\,\text{d}r + \int_{R_{\mathrm{c}}}^{R}r^{d-1}\frac{\partial c_{\mathrm{s}}}{\partial t}\,\text{d}r = D_{\mathrm{c}}\int_{0}^{R_{\mathrm{c}}}\frac{\partial}{\partial r}\left(r^{d-1}\frac{\partial c_{\mathrm{c}}}{\partial r}\right)\,\text{d}r + D_{\mathrm{s}}\int_{R_{\mathrm{c}}}^{R}\frac{\partial}{\partial r}\left(r^{d-1}\frac{\partial c_{\mathrm{s}}}{\partial r}\right)\,\text{d}r - k_{\mathrm{s}}\int_{R_{\mathrm{c}}}^{R}r^{d-1}c_{\mathrm{s}}\,\text{d}r,
\end{gather*}
due to radial symmetry. Reversing the order of integration and differentiation on the left, evaluating the first two integrals on the right and using the second interface condition (\ref{eq:model_cs_int}) yields:
\begin{gather*}
\frac{\text{d}}{\text{d}t}\left[\int_{0}^{R_{\mathrm{c}}}r^{d-1} c_{\mathrm{c}}\,\text{d}r + \int_{R_{\mathrm{c}}}^{R}r^{d-1}c_{\mathrm{s}}\,\text{d}r\right] = R^{d-1}\left[D_{\mathrm{s}}\frac{\partial c_{\mathrm{s}}}{\partial r}(R,t)\right] - k_{\mathrm{s}}\int_{R_{\mathrm{c}}}^{R}r^{d-1}c_{\mathrm{s}}\,\text{d}r,
\end{gather*}
and hence:
\begin{gather*}
R^{d-1}\left[-D_{\mathrm{s}}\frac{\partial c_{\mathrm{s}}}{\partial r}(R,t)\right] = -\frac{\text{d}}{\text{d}t}\left[\int_{0}^{R_{\mathrm{c}}}r^{d-1} c_{\mathrm{c}}\,\text{d}r + \int_{R_{\mathrm{c}}}^{R}r^{d-1}c_{\mathrm{s}}\,\text{d}r\right] - k_{\mathrm{s}}\int_{R_{\mathrm{c}}}^{R}r^{d-1}c_{\mathrm{s}}\,\text{d}r.
\end{gather*}
Integrating this last equation over the time interval $[0,\infty)$, using the initial condition (\ref{eq:model_cs_ic}) and the fact that both $c_{\mathrm{c}}(r,t)$ and $c_{\mathrm{s}}(r,t)$ tend to zero at infinite time yields:
\begin{gather*}
R^{d-1}\int_{0}^{\infty}\left[-D_{\mathrm{s}}\frac{\partial c_{\mathrm{s}}}{\partial r}(R,t)\right]\,\text{d}t = \int_{0}^{R_{\mathrm{c}}}r^{d-1}c_{0}\,\text{d}r - k_{\mathrm{s}}\int_{R_{\mathrm{c}}}^{R}r^{d-1}\mathcal{C}_{\mathrm{s}}(r)\,\text{d}r,
\end{gather*}
where we have set $\mathcal{C}_{\mathrm{s}}(r) := \int_{0}^{\infty} c_{\mathrm{s}}(r,t)\,\text{d}t$. Dividing both sides of this last equation by $\int_{0}^{R_{\mathrm{c}}}r^{d-1}c_{0}\,\text{d}r = (R_{\mathrm{c}}^{d}c_{0})/d$ yields the desired result:
\begin{gather*}
\frac{dR^{d-1}}{c_{0}R_{\mathrm{c}}^{d}}\int_{0}^{\infty}-D_{\mathrm{s}}\frac{\partial c_{\mathrm{s}}}{\partial r}(R,t)\,\text{d}t = 1 - \frac{k_{\mathrm{s}}d}{R_{\mathrm{c}}^{d}c_{0}}\int_{R_{\mathrm{c}}}^{R}r^{d-1}\mathcal{C}_{\mathrm{s}}(r)\,\text{d}r.
\end{gather*}
To complete the demonstration, we need to show that $\mathcal{C}_{\mathrm{s}}(r)$ satisfies the boundary value problem (\ref{eq:bvp_cs_ode1})--(\ref{eq:bvp_cs_bc}). The differential equations (\ref{eq:bvp_cs_ode1})--(\ref{eq:bvp_cs_ode2}) are obtained by integrating the reaction-diffusion equations (\ref{eq:model_cs_pde1})--(\ref{eq:model_cs_pde2}) over the time interval $[0,\infty)$, using the initial condition (\ref{eq:model_cs_ic}) and the fact that both $c_{\mathrm{c}}(r,t)$ and $c_{\mathrm{s}}(r,t)$ tend to zero at infinite time:
\begin{gather*}
-c_{0} = \int_{0}^{\infty}\frac{D}{r^{d-1}}\frac{\partial}{\partial r}\left(r^{d-1}\frac{\partial c_{\mathrm{c}}}{\partial r}\right)\,\text{d}t,\qquad 0 < r < R_{\mathrm{c}},\\
0 = \int_{0}^{\infty}\frac{D}{r^{d-1}}\frac{\partial}{\partial r}\left(r^{d-1}\frac{\partial c_{\mathrm{s}}}{\partial r}\right)\,\text{d}t - k_{\mathrm{s}}\int_{0}^{\infty}c_{\mathrm{s}}(r,t)\,\text{d}t,\qquad R_{\mathrm{c}} < r < R,
\end{gather*}
which simplify to equations (\ref{eq:bvp_cs_ode1})--(\ref{eq:bvp_cs_ode2}) when reversing the order of differentiation and integration in the second term on the right of both equations and introducing $\mathcal{C}_{\mathrm{c}}(r) := \int_{0}^{\infty} c_{\mathrm{c}}(r,t)\,\text{d}t$ and $\mathcal{C}_{\mathrm{s}}(r): = \int_{0}^{\infty} c_{\mathrm{s}}(r,t)\,\text{d}t$. Finally, the interface and boundary conditions (\ref{eq:bvp_cs_int})--(\ref{eq:bvp_cs_bc}) are obtained by integrating the interface and boundary conditions (\ref{eq:model_cs_int})--(\ref{eq:model_cs_bc}) over the time interval $[0,\infty)$ \cite{carr_2019}:
\begin{gather*}
\begin{cases} \displaystyle\int_{0}^{\infty}\frac{\partial c_{\mathrm{c}}}{\partial r}(0,t)\,\text{d}t = 0, & \text{if $d=1$},\\[0.4cm]
\text{$\displaystyle\int_{0}^{\infty}c_{\mathrm{c}}(0,t)\,\text{d}t$ finite}, & \text{if $d=2,3$},
\end{cases}
\qquad
\begin{cases} \displaystyle\int_{0}^{\infty} c_{\mathrm{s}}(R,t)\,\text{d}t = 0, & \text{if fully-permeable},\\[0.4cm]
\displaystyle-D\int_{0}^{\infty}\frac{\partial c_{\mathrm{s}}}{\partial r}(R,t)\,\text{d}t = P\int_{0}^{\infty} c_{\mathrm{s}}(R,t)\,\text{d}t, & \text{if semi-permeable},\\
\end{cases}
\end{gather*}
and reversing the order of differentiation and integration as appropriate.

\footnotesize
\setlength{\bibsep}{1pt plus 0.3ex}

\end{document}